\documentclass[prl,twocolumn,superscriptaddress,amsmath,amssymb,showpacs,floatfix,preprintnumbers]{revtex4}
\usepackage{amsmath}
\usepackage{graphicx}
\usepackage{dcolumn}
\usepackage{color}
\DeclareGraphicsExtensions{.png .jpg .pdf}

\begin{document}

\title{Plasmons in bias-induced topological phase transition in black phosphorus}

\author{D. J. P. de Sousa}\email{duarte.j@fisica.ufc.br}
\author{L. S. R. Cavalcante}\email{lucaskvalcante@fisica.ufc.br}
\author{Andrey Chaves}\email{andrey@fisica.ufc.br}
\author{J. Milton Pereira Jr.}\email{pereira@fisica.ufc.br}
\affiliation{Departamento de F\'isica, Universidade Federal do Cear\'a, Caixa Postal 6030, Campus do Pici, 60455-900 Fortaleza, Cear\'a, Brazil}
\author{Tony Low}\email{tlow@umn.edu}
\affiliation{Department of Electrical and Computer Engineering, University of Minnesota, Minneapolis, Minnesota 55455, USA}

\date{ \today }

\begin{abstract}
We investigate the plasmons in bilayer black phosphorus (BP) with bias-driven formation of Dirac cones, by developing an effective two-band Hamiltonian that captures this electronic transition with great accuracy. We show that the appearance of the Dirac cones lead to additional linearly dispersing acoustic plasmon mode, in conjunction to the conventional plasmon. In addition, the change in the Fermi surface topology from a disc to ring or dual pockets also modifies the dielectric loss.
\end{abstract}


\maketitle

One of the most remarkable features of the recently studied bi-dimensional (2D) materials is the possibility of controlling their electronic properties by means of external electric fields\cite{ref1, ref2, DOI:10.1103/PhysRevLett.99.216802}. Free carriers can be induced through chemical doping or electrical gating with great ease due to their 2D nature. Fields applied perpendicularly to the material plane can also modify the band structure of the system, as in the case of bilayer graphene, where it opens a gap in the electronic spectrum\cite{DOI:10.1103/PhysRevLett.99.216802}. Similarly, the electric field can also strongly modify the spectrum of BP\cite{Ye, Yuanbo}, which in bulk is a narrow gap semiconductor\cite{DOI:10.1021/nn501226z}. In contrast to graphene, in the case of few-layer BP, it has been shown that the application of an electric field would close the band gap, followed by the creation of two Dirac cones\cite{DOI:10.1103/PhysRevLett.119.226801,DOI:10.1038/srep11699,DOI:10.1021/acs.nanolett.5b04106}. These properties were recently investigated both theoretically\cite{DOI:10.1021/acs.nanolett.5b04106,DOI:10.1038/srep11699} and experimentally\cite{DOI:10.1103/PhysRevLett.119.226801,DOI:10.1103/PhysRevB.97.045143}.

The anisotropic optical properties of BP also present interesting opportunities for the study of anisotropic exciton and plasmon polaritons\cite{ref11, ref12}, hyperbolic plasmons\cite{ref13}, polarization sensitive optoelectronics\cite{ref14, ref15}, and atomic waveplates\cite{ref16}. The ability to modify the electronic band structure, and consequently its anisotropy, through electric field would enable new means of tuning in-plane optical birefringence on-demand.
	
In this work, we investigate the bias-driven formation of Dirac cones and how this transition affects the anisotropy and dispersion of plasmons, and its nonlocal dielectric loss. Our results show that as the field increases, the formation of two cones in the energy spectrum with distinct dispersions along the in-plane crystal axes, leads to the appearance of acoustic-like plasmons modes and opens up additional channel to dielectric loss.

The behavior of low energy electrons and holes in multilayer is well described by the tight-binding (TB) model proposed by Rudenko \textit{et al}.\cite{Rudenko}. The multilayers are assumed to be stacked in an AB configuration, which is most energetically favorable\cite{http://dx.doi.org/10.1116/1.4926753}. Such model, which is very accurate in describing multilayer BP and has successfully been used in several recent works\cite{Boundary condition,10.1103/PhysRevB.96.155427}, becomes computationally expensive for simulating large area systems in the presence of external perturbations. Thus, several approximations within the context of the TB model have been proposed for studying BP systems, some of which exhibit excellent agreement with previous models\cite{Boundary condition,10.1103/PhysRevB.96.155427,Milton}. Although some of these approaches are suitable for describing the behavior of gated multilayer systems\cite{Milton}, they are still intractable for an analytical approach. To the best of our knowledge, there is no continuum model available for describing accurately the energy bands under large applied bias, especially in the regime where the gap closes and with the emergence of Dirac spectrum\cite{DOI:10.1038/srep11699}. Therefore, we first propose an effective Hamiltonian that describes the low energy bands of a biased bilayer of BP with good accuracy, even at the vicinity of the critical fields corresponding to the closure of the gap, where it still shows excellent agreement with the tight-binding (TB) approach of Ref.~[\onlinecite{Rudenko}]. 

\begin{figure}[t]
\centerline{\includegraphics[width = \linewidth]{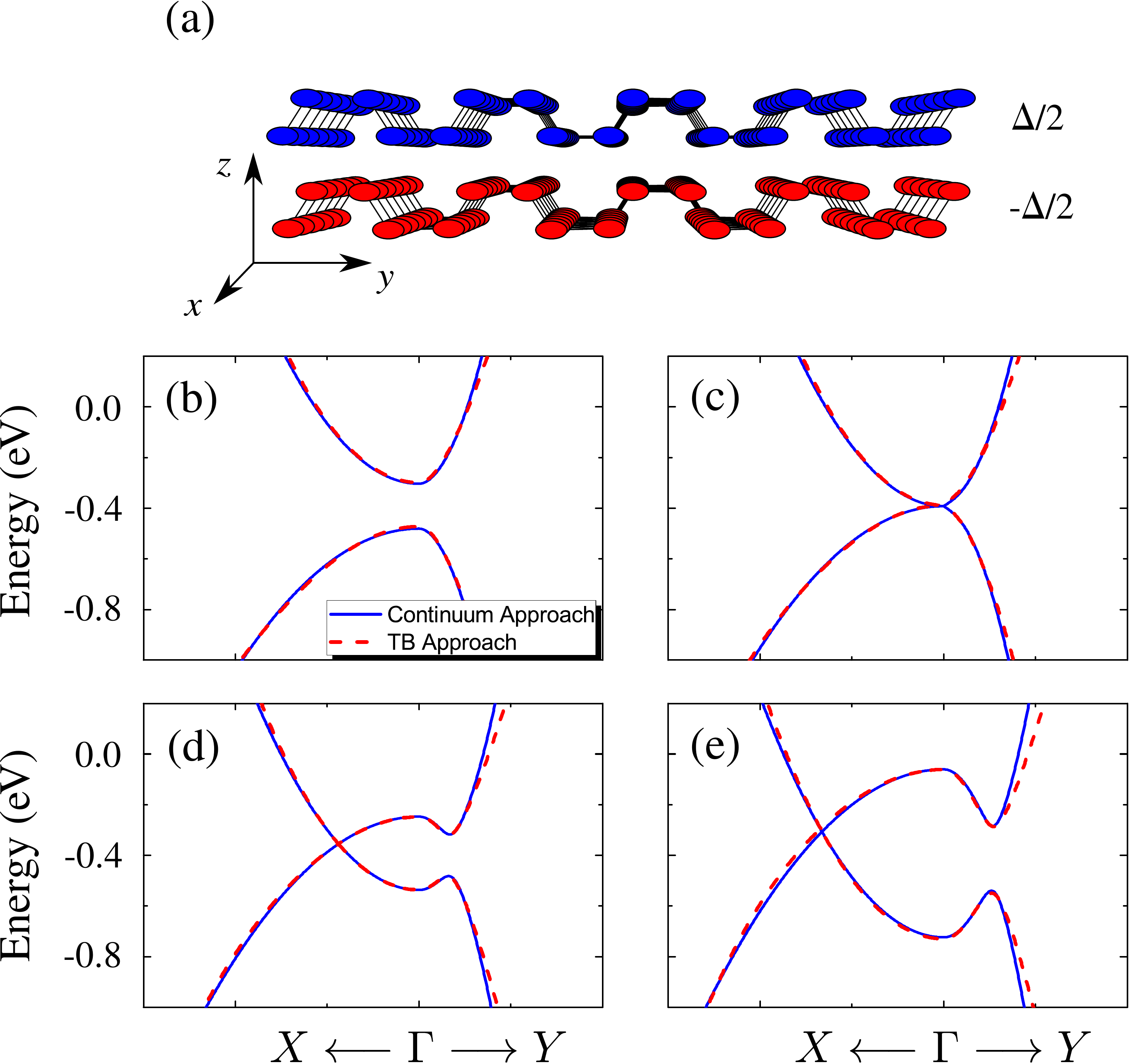}}
\caption{(Color online) (a) Bilayer BP with applied
bias. The onsite energy of all phosphorus atoms at the top
layer, represented by the blue atoms, is assumed to be $\Delta/2$
whereas the onsite energy of the atoms at the bottom layer,
represented by the red atoms, is assumed to be $−\Delta/2$. Energy band for different bias : (b) $\Delta$ = 1.5 eV, (c) $\Delta$ = 1.69 eV, (d) $\Delta$ = 2.0 eV and (e) $\Delta$ = 2.4 eV. The dashed lines represent the results from the TB model whereas the solid blue lines represent the results obained from Hamiltonian (\ref{Eq4}).}
\label{Fig2}
\end{figure}

For the sake of simplicity, we study the system sketched in Fig.~\ref{Fig2}(a) : Bilayer BP with an applied bias $\Delta$, in the out-of-plane direction, such that the on-site energy at the top (bottom) monolayer, represented by the blue (red) atoms, is $\Delta$/2 ($-\Delta$/2). Since we are concerned with the physics at the vicinity of the gap closing at critical bias $\Delta_c \approx 1.69$ eV, we propose the following two band Hamiltonian for the purpose of the calculation of the spectrum [See Suplementary Information]
\begin{equation}\mathcal{H} = \left(
\begin{array}{cc}
u_0 + \eta_x k_ {x}^{2} + \eta_y k_{y}^{2} & f + \gamma_x k_ {x}^{2} + \gamma_y k_{y}^{2} + i\chi k_{y} \\
f + \gamma_x k_ {x}^{2} + \gamma_y k_{y}^{2} - i\chi k_{y} & u_0 + \eta_x k_ {x}^{2} + \eta_y k_{y}^{2} 
\end{array}\right),
\label{Eq4}
\end{equation}
where $f = (\Delta_c - \Delta)/\delta$, with the following values of the coefficients : $u_0 = -0.392$ eV, $\delta = 2.14$, $\eta_x = 0.778$ eV $\cdot$ \AA$^2$, $\eta_y = -1.139$ eV $\cdot$ \AA$^2$, $\gamma_x = 3.02$ eV $\cdot$ \AA$^2$, $\gamma_y = 17.56$ eV $\cdot$ \AA$^2$, $\chi = 0.95$ eV$\cdot$ \AA .	

Equation~(\ref{Eq4}) has the general form of the \textit{n}-th band sub-Hamiltonian for the N-layer BP system in the absence of sublattice symmetry breaking in each layer\cite{10.1103/PhysRevB.96.155427}. In this work, we incorporate the effect of the bias $\Delta$ through $f = f(\Delta)$. It is worth mentioning that the same Hamiltonian is not limited to bilayer systems. In fact, the merging of the lowest energy bands of multilayer BP system is known to be similar to the bilayer case. For a system with an arbitrary number of layers, one would have to find the values of the coefficients suitable for the different cases, whereas the structure of the Hamiltonian would remain the same. 

The comparison between the energy bands obtained from Eq.~(\ref{Eq4}) and the full TB Hamiltonian of Ref.~[\onlinecite{Rudenko}] is presented in Fig.~\ref{Fig2}(b)-(e) for different biases. It is seen that the merging bands and consequent formation of Dirac spectrum, resulting from continuum approximated Hamiltonian, represented by the solid blue curves, show good agreement with the results obtained by the TB model, represented by the dashed red curves. We have found that the continuum description is accurate in the range $1.2$ eV $< \Delta < 2.5$ eV. 
	
For biases smaller than $\Delta_c \approx 1.69$ eV ($\Delta = 1.5$ eV in Fig.~\ref{Fig2}(b)), the typical anisotropic energy bands for electrons and holes are observed. This corresponds to the case where there is still an energy gap in the system. At the critical bias $\Delta_c$ in Fig.~\ref{Fig2}(c), the valence and conduction bands touch while maintaining a similar ratio between the effective masses along the zigzag ($\Gamma \rightarrow X$) and armchair ($\Gamma \rightarrow Y$) direction. However, for $\Delta > \Delta_c$, one sees an inversion of the energy bands which leads to the formation of the Dirac spectrum (see Figs.~\ref{Fig2}(d) and (e)). This transition is more pronounced for higher fields, as one can see in Fig.~\ref{Fig2}(e).

Taking into account only the off-diagonal elements of Hamiltonian (\ref{Eq4}), without loss of generality, one can write the energy bands as
$\epsilon_{\pm} =\pm \sqrt{(f + \gamma_x k_x^{2} + \gamma_y k_y^{2})^{2} + (\chi k_y)^{2}}$. Such description allows us to consider the dispersion along $\Gamma \rightarrow X$ more easily. Considering $\Delta > \Delta_c$, i.e. the gapless regime, we find $\epsilon_{\pm} = \pm(f + \gamma_x k_x^{2})$ at $k_y =0$. The bands intersect at two points with momentum given by $\Lambda_x = \pm \sqrt{-f/\gamma_x}$, which are the locations of the gapless Dirac points in momentum space. Therefore, we find that the $k$ splitting of the Dirac cones depends on the bias as $2\sqrt{-(\Delta_c - \Delta)/\delta\gamma_x}$, which is in agreement with recent experimental measurements\cite{DOI:10.1103/PhysRevLett.119.226801}.

The proposed Hamiltonian describes properly the physics at the vicinity of the $\Lambda_x$ points. In fact, we can expand the elements of the continuum Hamiltonian around $\Lambda_x$ by taking $k_x = \Lambda_x - q_x$ and consider only linear contributions in momenta: 
\begin{equation}
\mathcal{H} = \left(
\begin{array}{cc}
0 & -2\gamma_x\Lambda_x q_x + i\chi q_{y} \\
-2\gamma_x\Lambda_x q_x - i\chi q_{y} & 0
\end{array}\right),
\label{Eq6}
\end{equation}
since $f + \gamma_x \Lambda_ {x}^{2} = 0$. The above Hamiltonian is an anisotropic version of the massless Dirac Hamiltonian for 2D electrons with velocities $v_x = 2\gamma_x \Lambda_x/\hbar$ and $v_y = \chi/\hbar$ along the $x$ and $y$ directions, respectively.

The Hamiltonian~(\ref{Eq4}) has a similar structure as the universal Hamiltonian proposed by Montambaux \textit{et al.} to describe the formation of Dirac points in the electronic spectrum\cite{10.1103/PhysRevB.80.153412}. Despite the subtle differences, we expect to observe similar emerging features. Indeed, Montambaux's analysis of the dependence of the Landau levels as a function of the merging parameter is an excellent qualitative description of the results found by Pereira and Katsnelson for the biased bilayer BP case\cite{Milton}. However, Montambaux's Hamiltonian fails to describe correctly BP systems as it finds a $B^{2/3}$ dependence of the Landau levels, whereas the TB spectrum was found to be linear in the magnetic field\cite{10.1021/acs.nanolett.7b03954}, and do not take into account the transition from a disc to a ring or dual pockets topology of the 2D Fermi surface as the gap closes. Both characteristics emerge naturally from Hamiltonian~(\ref{Eq4})\cite{Milton,10.1103/PhysRevB.96.155427}.
\begin{figure}[t]
\centerline{\includegraphics[width = \linewidth]{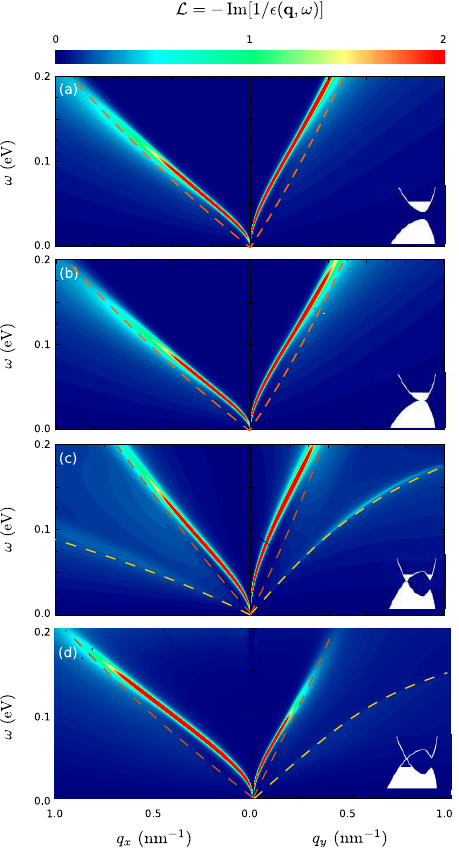}}
\caption{(Color online) Intensity map of the energy loss function considering an electron doping of $1\times 10^{13}$ cm$^{-2}$ in (a) and (b) for biases $\Delta = 1.5$ eV and $1.69$ eV, respectively, and of $2\times 10^{13}$ cm$^{-2}$ electron and hole doping in (c) and (d), respectively, considering $\Delta = 2.3$ eV. The insets show the band diagrams and the Fermi level along the x and y directions corresponding to each case. Dashed lines represent the edges of the electron-hole excitation continua.}
\label{Fig3}
\end{figure} 
 
In the following, we apply Eq.~(\ref{Eq4}) to study the behavior of plasmons at the vicinity of the band transition in biased bilayer BP. A common approach to describe the plasmonic properties of materials is through the calculation of the dielectric function of the system $\epsilon(\textbf{q}, \omega)$, from which one can directly determine the energy loss function, defined as $\mathcal{L}(\textbf{q}, \omega) = -\operatorname{Im}[1/\epsilon(\textbf{q}, \omega)]$\cite{Nuno, Giuliani}.

The dielectric function in the random phase approximation (RPA) is given by
\begin{equation}
\epsilon(\textbf{q}, \omega) = 1 + v_{c}(q)\Pi (\textbf{q}, \omega),
\label{eq11}
\end{equation}
where $v_c (q) = 2\pi e^2 /\kappa q$ is the 2D Coulomb interaction and $\kappa$ describes the effective dielectric constant of the two half-spaces mediums surrounding the plane, where we assume $\kappa  \approx 2.5$ corresponding to a common SiO$_2$ substrate and air. The pair-bubble diagram contribution is 
\begin{equation}
\Pi (\textbf{q}, \omega) = -2\sum_{j,j^{'}}\int \frac{d^2 k}{(2\pi)^2} \frac{(f_{j,\textbf{k}}-f_{j^{'},\textbf{k} + \textbf{q}})|\langle\Phi_{j,\textbf{k}}|\Phi_{j^{'},\textbf{k}+\textbf{q}}\rangle|^2}{\hbar\omega + \epsilon_{j,\textbf{k}}-\epsilon_{j^{'}, \textbf{k}+\textbf{q}} + i\eta},
\label{eq12}
\end{equation}
where the Fermi-Dirac distribution is $f_{j,\textbf{k}} = [\exp(\beta (\epsilon_{j\textbf{k}} - \mu)) + 1]^{-1}$ with $\beta = 1/k_B T$ given in terms of the temperature, which we will assume to be $T = 300$ K in our calculations. Additionally, we assume a phenomenological broadening factor to be $\eta = 1$ meV.

Figure~\ref{Fig3} shows the energy loss function $\mathcal{L}$ along the main crystallographic directions, $x$ and $y$, for several biases $\Delta$, where we have assumed a constant electron doping of $1\times 10^{13}$ cm$^{-2}$ in (a) and (b). The dashed lines represent the edges of the electron-hole excitation continua. In Fig.~\ref{Fig3}(a), we selected $\Delta = 1.5$ eV, which corresponds to the regime where there is still an energy gap in the system, as shown by the band diagram inset. We obtain a spectrum similar to that of monolayer BP of Ref.~[\onlinecite{10.1103/PhysRevLett.113.106802}], where a clear anisotropic dispersion leads to higher dispersive plasmons along the $y$ direction (i.e. armchair direction). We observe no significant change as we approach the critical value $\Delta = \Delta_c$ [See Fig.~\ref{Fig3}(b)]. This is due to the fact that the shape of the conduction and valence bands are still very similar to the corresponding ones for $\Delta < \Delta_c$. The situation is different for $\Delta > \Delta_c$, where the valence and conduction bands undergo a phase transition which causes the formation of Dirac-like spectra, as previously discussed. Figure~\ref{Fig3}(c) shows the results corresponding to an applied bias of $\Delta = 2.3$ eV and an electron doping of $2\times 10^{13}$ cm$^{-2}$, where the loss function is now significantly different from the previous cases due to the change in the topology of the bands. Also, the increase in doping can modify the single particle transition continuum, i.e. the intra and inter band Landau damping, leading to different damping frequencies. 

Moreover, we also notice the appearance of weaker peaks in the loss function along both crystallographic directions for $\Delta > \Delta_c$, seen in Fig.~\ref{Fig3}(c). These additional modes are less energetic than the well known plasmon modes and present linear dispersions for small momenta, i.e. acoustic modes. Notice that our model does not allow the study of effects that can arise due to the presence of separate layers, since we are using an effective monolayer-type Hamiltonian with modified parameters. Therefore, we cannot attribute the existence of these acoustic-like plasmon modes to the out-of-phase oscillations of the charge densities in the separate BP layers \cite{ref27, ref28,ref29}. The typical 2D plasmon modes for $\Delta > \Delta_c$ with conventional $\propto \sqrt{q}$ dispersion for small momenta is due to the in-phase oscillations of charge carriers with different character (i.e. effective mass), whereas the acoustic plasmon modes with $\propto q$ dispersion are due to the out-of-phase fluctuations of these electrons. Acoustic plasmon modes due to multi-carrier environment has been predicted in various 2D systems, e.g. multilayer black phosphorus \cite{ref30}, double-layer graphene \cite{ref29}, as well as in transition metal dichalcogenides (TMDs) \cite{ref31}. 

Since the system is not electron-hole symmetric, we do not expect the plasmons character to be similar in the hole doped case. Figure.~\ref{Fig3}(d) shows the plasmon with similar hole doping instead. The inset reveals a prominent Dirac dispersion along the zigzag, but is gapped out along the armchair. Since plasmons is due to the intraband processes, a substantial part of the Drude weight is transferred to the plasmon dispersing along zigzag. Hence, we observed a more prominent plasmon mode along zigzag which damped out at much higher energy than the mode along armchair. This is in contrary to the electron doped situation depicted in Fig.~\ref{Fig3}(c). In general, the plasmons as depicted in the different cases in Fig.~\ref{Fig3} would also exhibit different scaling with carrier densities \cite{10.1103/PhysRevLett.113.106802}

\begin{figure}[t]
\centerline{\includegraphics[width = \linewidth]{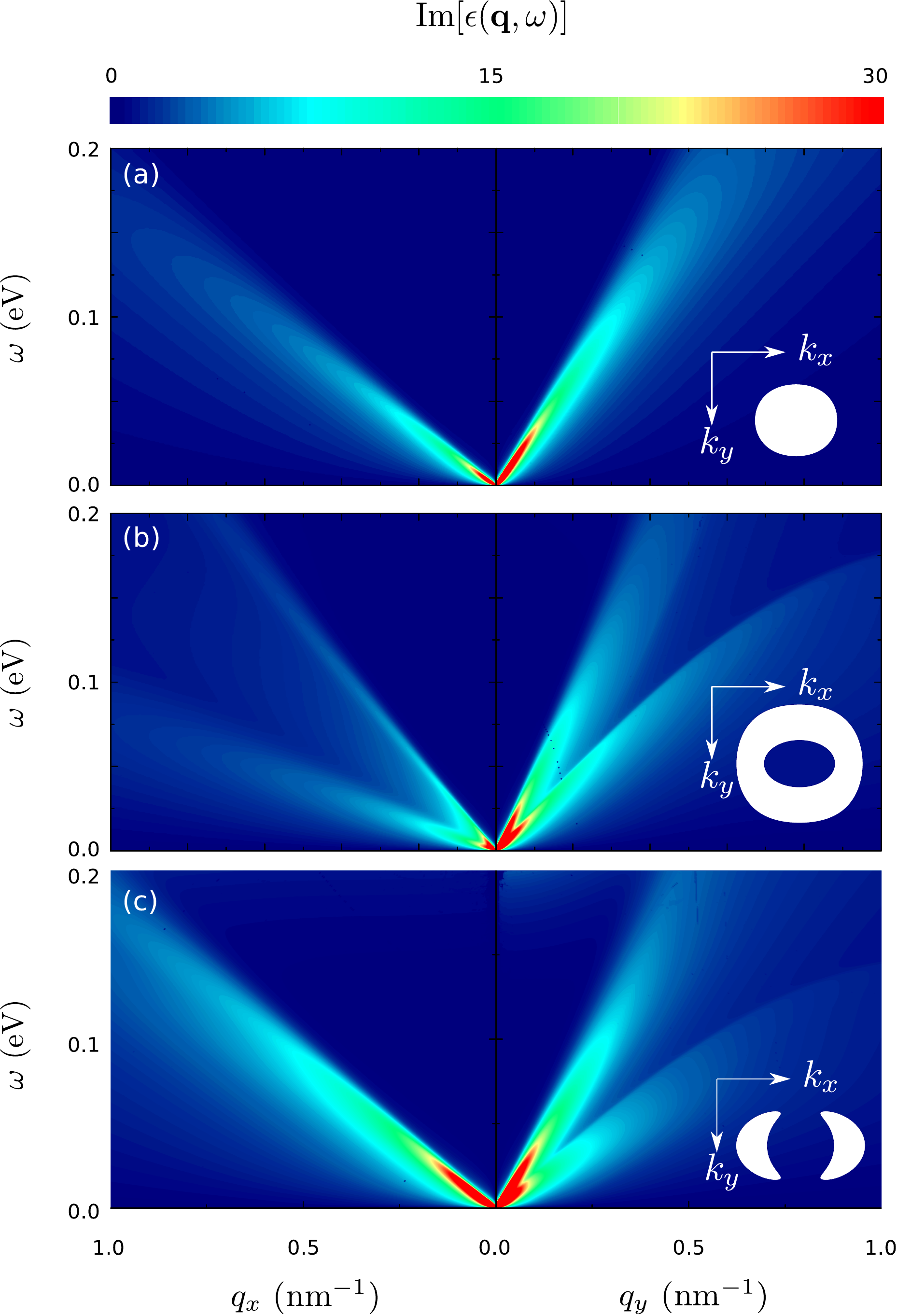}}
\caption{(Color online) Intensity map of the imaginary part of the dielectric function considering an electron doping of (a) $1\times 10^{13}$ cm$^{-2}$ and $\Delta = 1.69$ and of $2\times 10^{13}$ cm$^{-2}$ electron and hole doping in (b) and (c), respectively, considering $\Delta = 2.3$ eV. Insets show the 2D Fermi surfaces for each case.}
\label{Fig4}
\end{figure}

The optical absorption of freestanding 2D materials is related to the losses of its dielectric function, i.e. $\operatorname{Im}\epsilon(\textbf{q},\omega=0)$. In the non-local limit, this  is  related to its optical conductivity via the charge-current continuity equation $\operatorname{Im}\epsilon(\textbf{q},\omega) = q/2\epsilon_0\omega \operatorname{Re}\sigma(\textbf{q},\omega)$. The imaginary part of $\epsilon (\textbf{q}, \omega)$ for the cases presented in Fig.~\ref{Fig3} are shown in Fig.~\ref{Fig4}. This quantity tracks the losses in the system due to single particle transitions, i.e. the electron-hole excitation continuum. The intensity map of Fig.~\ref{Fig4}(a) shows the anisotropic behavior of $\operatorname{Im}\epsilon(\textbf{q},\omega)$ for $\Delta = \Delta_c$, which resemble those of monolayer BP systems\cite{10.1103/PhysRevLett.113.106802} as one would expect. However, additional damping channel in the spectrum is observed when $\Delta > \Delta_c$, as shown in Fig.~\ref{Fig4}(b). As one can see, the contributions to the losses reside in two distinct regions along each crystallographic direction. The introduction of an additional damping channel is due to the fact that, for the assumed electron dopings, the 2D Fermi surface has now the topology of a ring, instead of the Fermi disc topology in Fig.~\ref{Fig4}(a) as shown in the insets. Similar topological transition was also studied in the case of biased bilayer graphene \cite{ref32}. For biases smaller than $\Delta_c$, a given momentum transfer $q$ excites an electron out of the Fermi disc, creating a hole inside the 2D Fermi surface. Interestingly, if the 2D Fermi surface has a ring topology, the electron could also be excited to the region inside of the ring. In this case, if the effective mass of the electron is different in this region, a second single particle phase space appears with different dispersion. The appearance of this second dispersion is also presented by dashed curves in Fig.~\ref{Fig3}. In the hole doped case as shown in Fig.~\ref{Fig4}(c), the Fermi surface acquires a very different topology with two Fermi pockets unlike its electron doped counterpart. In the hole case, the Dirac dispersion along the zigzag is more symmetric, hence its loss spectrum reveals only a single branch. With an electrical bias in BP, we have shown that topological transition in its electronic spectrum results in nontrivial modifications to its dielectric loss spectrum.

In summary, we have proposed a two-band Hamiltonian to describe the electronic properties of multilayer BP systems under the influence of an external bias. We have shown that such Hamiltonian is suitable for describing the topological phase transition of the lower energy bands of BP systems with great accuracy. We also have studied the behavior of the energy loss function in BP systems at the vicinity of the critical bias $\Delta_c$. In particular, we have shown the appearance of acoustic-like plasmon modes for $\Delta > \Delta_c$. Our results suggest that these modes are due to the unique warping of the energy bands. We have also discussed the influence of the electrical bias on the topology of the Fermi surface and its dielectric loss. Our studies should have important consequences on the optoelectronics of BP systems as it provides a way to control light-matter interaction properties through electrical biasing.

\emph{Acknowledgement.} This work was financially supported by the CAPES foundation. T.L acknowledges support from the National Science Foundation under Grant No. NSF/EFRI-1741660.

\end{document}